\documentstyle[12pt,fleqn]{article}

\newcommand{\AmS}{{\protect\the\textfont2
  A\kern-.1667em\lower.5ex\hbox{M}\kern-.125emS}}

\begin{document}
\begin{center}
{\Large{\bf Isoscalar Roper Excitation in $p(\alpha, \alpha')$
       Reactions in the $10 - 15$ $GeV$ Region }}

\vspace{1.0truecm}

{S. Hirenzaki$^{a,}$
\footnote { JSPS research fellow }
, E. Oset$^a$, P. Fern\'andez de C\'ordoba$^b$ }

\vspace{1.0truecm}

{$^a$ Departamento de F\'{\i}sica Te\'orica and IFIC, Centro Mixto
\\Universidad de Valencia - CSIC,
46100 Burjassot (Valencia) Spain.\\}
{$^b$ Departamento de Matem\'atica Aplicada, Universidad Polit\'ecnica de
Valencia,
\\46022 Valencia, Spain.\\}

\vspace{2.0truecm}

\end{center}

\begin{abstract}
Recent experiments at Saturne at $4$ $GeV$ showed that the $(\alpha,\alpha')$
reaction on the proton shows two distinctive peaks, which were associated to
$\Delta$ projectile excitation and Roper target excitation.  A subsequent
theoretical analysis has shown that this picture is qualitatively correct
but there are important interference effects between the two mechanisms.
Futhermore, at this energy the ratio of strengths for the Roper and $\Delta$
peak is about $1/4$.  In the present paper we show that by going to the
$10 - 15$ $GeV$ region the interference effects become negligible, the
signal for the Roper excitation is increased by more than an order of
magnitude and the ratio of cross sections at the peaks for Roper
and $\Delta$ excitation becomes of the order of unity, thus making
this range of energies ideal for studies of isoscalar Roper excitation.
\end{abstract}

\newpage

\setlength{\baselineskip}{8mm}

The $(\alpha,\alpha')$ reaction on proton targets at kinetic energy
$T_{\alpha}=4.2$ $GeV$ was studied at SATURNE \cite{Morsch92} and two
distinctive peaks were identified (see Fig. 1), which were associated
to $\Delta$ excitation in the $\alpha$ projectile and Roper excitation
in the proton target (see Fig. 2).  In a recent theoretical analysis
we showed that the two mechanisms of Fig. 2 were dominant in the reaction and
that other possible mechanisms, like Roper excitation on the projectile or
two $\Delta$ excitation, were negligible \cite{HirenSub}.  However,
it was found that the interference between the two mechanisms in Fig. 2
was appreciable and it was important to consider for a proper analysis
of the data and the excitation of the isoscalar $NN \rightarrow NN^*$
transition amplitude.

In Fig. 1 one can see the results for the projectile $\Delta$ excitation,
Roper target excitation and interference.  One observes there that the
interference term is large and that the strength of the Roper is about
$1/4$ of the strength of the $\Delta$ excitation at their peaks.

It would be interesting to have other experiments which magnified
the strength of Roper excitation, both in absolute terms and
relative to the $\Delta$ and if possible diminished the interference
term, which makes a theoretical model necessary in order to separate the Roper
contribution. All these things are accomplished by performing the
$(\alpha,\alpha')$ reaction at higher energies, as we explain here.

We take the same model which was used in \cite{HirenSub} to analyse the
$(\alpha,\alpha')$ reaction at $4$ $GeV$.  The cross section for the processes

\begin{displaymath}
   \alpha + p \rightarrow \alpha + p + \pi^0
\end{displaymath}

\begin{equation}
   \alpha + p \rightarrow \alpha + n + \pi^+
\end{equation}

\noindent
is given by

\begin{displaymath}
   \frac{d^2{\sigma}}{dE_{\alpha'}d\Omega_{\alpha'}} =
   \frac{p_{\alpha'}}{(2{\pi})^5}
   \frac{M_{\alpha}^2 M^2}{\lambda^{1/2}(s,M^2,M_{\alpha}^2)}
   \int d^3 p_{\pi}
   \frac{1}{E_{N'} \omega_{\pi}}
\end{displaymath}

\begin{equation}
   \times \bar{\Sigma} \Sigma \vert T \vert ^2
   \delta (E_{\alpha} + E_N - E_{\alpha'} - E_{N'} - \omega_{\pi})  ,
\end{equation}

\noindent
where $\lambda (...)$ is the K\"allen function and $s$ the Mandelstam
variable for the initial $p - \alpha$ system.

By means of eq. (2) we can take into account the mechanisms of
$\Delta$
excitation in the projectile, Fig. 2a, and the Roper excitation in the
target, Fig. 2b, with the Roper decaying into a nucleon and a pion (which
accounts for about $ 65 \% $ of the $ N^* $ free width).
The contribution of the Roper decay into
$ \pi \pi N$ is also
accounted for in \cite{HirenSub} and
is included in the final results here but it does not interfere with the
amplitude of $ \Delta $ excitation in the projectile, since the final
states are different.

The $T$ matrix for the diagram 2a is evaluated taking into account
$\pi + \rho$ exchange together with the Landau Migdal induced correction.
For the diagram of Fig. 2b, which enforces the exchange of an isoscalar
object, we take an effective $"\sigma"$ exchange, which incorporates the
possible exchange of an $\omega$ meson and the effect of nuclear correlations.
The strength of this isoscalar exchange piece is determined by making a fit
to the experiment of ref. \cite{Morsch92}.  The expressions for the $\Delta$
and Roper terms and the interference can be seen in eq. (4) of ref.
\cite{Pedro95} and eqs. (4), (20) of ref. \cite{HirenSub}. Their reproduction
here is not necessary to understand the results.

We have evaluated the cross section for the $(\alpha,\alpha')$ reaction
on the proton, using the same model, for kinetic energies $10$ $GeV$
and $15$ $GeV$
of the $\alpha$ particle.  We show in Fig. 3 the results obtained for
$15$ $GeV$.  Those at $10$ $GeV$ are qualitatively similar but the Roper
and $\Delta$ peaks have a strength of about $ 4$
$[mb/(sr \cdot MeV)] $.  In Fig. 3
we show the results of the Roper excitation (with decay of the Roper
into $\pi N$), those of the $\Delta$ projectile excitation, their
interference and the sum, which includes also the contribution of the
$N^* \rightarrow \pi \pi N$ decay (with the distortion of the two pions by
the $^4 He$ nucleus which must remain unbroken).  Comparison of Fig. 3
with Fig. 1 shows the welcome feature of the $15$ $GeV$ reaction:

\noindent
i) The cross section for the Roper excitation is increased by more
than an order of magnitude with regard to the one at $4$ $GeV$.

\noindent
ii) The strength of the Roper and $\Delta$ peaks is similar, while
at $4$ $GeV$ the former had a strength of about $1/4$ of the latter.

\noindent
iii) The interference term is practically negligible compared to the
Roper contribution.  This is in contrast with the $4$ $GeV$ case where
the strength of these two terms was similar.

A situation like the one in Fig. 3 makes experimentally much easier the
extraction of information on the properties of Roper excitation by an isoscalar
source.  Such experiments can be easily implemented in the
Synchrophasotron of Dubna which accelerates nuclei up to $T_{kin}
\simeq 4$ $GeV/A $ and in the new superconductive synchrotron,
the Nuclotron, which accelerates nuclei up to $T_{kin} \simeq 6$ $GeV/A$
\cite{Strok95}.  In fact in a related experiment carried out at Dubna
on the $C(d,d')X$ reaction \cite{Ableev83} at $8.9$ $GeV/c$, a reanalysis
of the data in terms of $M_X$ calculated for the $p(d,d')$ kinematics
shows a clear peak around the Roper mass \cite{StrokPC}.

It is relatively easy to understand the features observed in the
results at $15$ $GeV$. In the first place the small interference.  It is
easy to see that as the energy of the beam increases it becomes
progressively more difficult to have the same kinematic configuration
of $\alpha, N, \pi$ in the final state for the two mechanism of Fig. 2.
Indeed, in the lab. system the pion coming from the decay of the Roper
in the mechanism of Fig. 2b will be distributed in a wide range of angles
( it would be isotropic for the $N^*$ decay at rest, but the effective
$\sigma$ brings some momentum along).  However, the pion coming from
the $\Delta$ decay in the mechanism of Fig. 2a will be directed in a very
narrow cone along the direction of motion of the $\alpha$ particle
in the frame where the initial proton is at rest.  The cone becomes
narrower as the $\alpha$ particle energy increases and, hence, the overlap
of the final state configurations in the two mechanisms of Fig. 2 (and
the interference term) becomes smaller as $T_{\alpha}$ increases.

In order to understand the change of strength of the Roper and $\Delta$
excitations and their relative weight we must look at another factor.
The reason in this case lies in the nucleus form factor which one has
in this reaction.

Indeed, in both the mechanisms of Fig. 2 the amplitude contains the
nuclear form factor \cite{Pedro95}

\begin{displaymath}
  F_{He}(\vec{k})  = \int d^3 r \rho_{He} (\vec{r})
               exp \left[ - \frac{1}{2} \int _{-\infty} ^{\infty} \sigma_{NN}
               \rho_{He} (\vec{b}, z') dz' \right]
               e^{i \vec{k} \cdot \vec{r} }
\end{displaymath}

\begin{equation}
               \times exp \left[ - \frac{i}{2} \int _{0} ^{\infty}
               \frac{1}{p_{\pi}} \Pi (p_{\pi}, \rho_{He} (\vec{r'} ) ) d \ell
               \right]  ,
\end{equation}

\noindent
where

\begin{displaymath}
   \vec{r'} = \vec{r} + \frac{\vec{p_{\pi}}}{\vert \vec{p_{\pi}} \vert} \ell ,
\end{displaymath}

\begin{equation}
  \vec{k} = \vec{p_{\alpha}} - \vec{p_{\alpha'}}  .
\end{equation}

\noindent
The momenta $\vec{p_{\alpha}}$, $\vec{p_{\alpha'}}$, $\vec{p_{\pi}}$ appearing
in eqs. (3), (4) are evaluated in the frame where the initial $\alpha$
particle is at rest.  In eq. (3) $\rho_{He} (\vec{r})$ is the Harmonic -
Oscillator density distribution of the
$\alpha$ particle, $\sigma_{NN}$ the nucleon-nucleon total cross section
and $\Pi (p_{\pi},\rho) / 2 \omega_{\pi}$ is the pion nuclear optical
potential, taken from ref. \cite{Nieves93} up to $T_{\pi} \simeq 250 MeV$ and
extrapolated at high energies when needed using the lowest order optical
potential \cite{HirenSub}.

The form factor of eq. (3) is the $^4 He$ nuclear form factor incorporating
the distortion of the proton and pion waves, both in the eikonal
approximation.  Now when $T_{\alpha}$ increases, for a same energy transfer
the momentum transfer is smaller.  Indeed in the forward direction of the
$\alpha'$ we have

\begin{displaymath}
 p_{\alpha} - p_{\alpha'} = \sqrt{E^2 _{\alpha} - M^2_{\alpha}}
                          - \sqrt{E^2 _{\alpha'}- M^2_{\alpha}}
\end{displaymath}

\begin{equation}
                          \simeq E_{\alpha} - E_{\alpha'}
                          -  \frac{M^2 _{\alpha}}{2 E_{\alpha}}
                          +  \frac{M^2 _{\alpha}}{2 E_{\alpha'}}
                 = (E_{\alpha} - E_{\alpha'})
                   \left( 1 + \frac{M^2 _{\alpha}}{2 E_{\alpha} E_{\alpha'}}
                   \right)  .
\end{equation}

\noindent
Hence, the invariant four momentum transfer squared will be

\begin{equation}
 -q^2 \simeq ( E_{\alpha} - E_{\alpha'} )^2
      \left( \left[ 1 + \frac{M_{\alpha}^2}{2 E_{\alpha} E_{\alpha'}}
             \right] ^2 - 1 \right)
\end{equation}

\noindent
which decreases as $E_{\alpha}$ increases.
The magnitude $ -q^2$ is equivalent to $\vec{k}^2$
in the Breit frame of the nucleus and is essentially also $\vec{k}^2$ of
eqs. (3), (4) for the $^4 He$ at rest.  Hence, we should expect an increase
of the
value of the nucleus form factor, $F_{He}(\vec{k})$ as $E_{\alpha}$ increases.
Futhermore the relative increase in the form factor (think in terms of the
undistorted $ exp[ - \vec{k}^2 / 4 \alpha ^2 ]$ form factor of $^4 He$
with $\alpha^2 = 0.76 fm^{-2}$ ) will
be bigger if the excitation energy $E_{\alpha} - E_{\alpha'}$ is bigger.
This is actually what we see in Fig. 4, where we plot the ratio of
$ \vert F_{He} (\vec{k}) \vert ^2 $ at two different $T_{\alpha}$ energies
, $10$ $GeV$ and $4.2$ $GeV$, and $15$ $GeV$ and $4.2$ $GeV$.

We observe in Fig. 4 that at $\omega = E_{\alpha} - E_{\alpha'}$ around
$200$ $MeV$, where the $\Delta$ peak appears, the increase of the form factor
is moderate.  However, at $\omega = 550$ $MeV$, where the Roper peak appears,
the ratio of form factors squared has a value of the order of four to five.
This factor is the one responsible for the relative increase of strength of the
Roper excitation versus the $\Delta$ excitation at $T_{\alpha}=15$ $GeV$
with respect to the experiment of \cite{Morsch92} at $4.2$ $GeV$.

The absolute increase in strength both for the Roper and $\Delta$
excitation can be traced back both to the form factor effect and the phase
space factor $p_{\alpha'}$ in the numerator of the cross section
formula of eq. (2).

One might think that performing more exclusive experiments, $i. e.$,
detecting a pion and a nucleon in coincidence and making a plot in terms of
the $\pi N $ invariant mass, one would magnify the Roper peak with respect to
a background. We have checked theoretically, within our model, that this
is not the case and the invariant mass distribution which one obtains resemble
very much the $\omega$ distributions of Fig. 1 and 3.

In our theoretical model we have neglected any dependence of the interaction
on the energy, since we have no elements to think that this might be
the case.  Obviously a certain energy dependence cannot be ruled out,
an information which would be provided by the same experiment and which
would be much useful to help construct microscopic models for the
interaction.

The results obtained here should encourage the implementation of the
experiments.  After decades of studies around the $\Delta$ region the
time has come to study in detail the properties of the next nucleon
excitation.  Quark models have difficulties to explain the properties
of the Roper \cite{Silv85}; the authors of ref. \cite{Morsch92} suggested
that the Roper could be interpreted as a monopole excitation of the nucleon
(breathing mode); the decay of the Roper into two pions in S-wave plays an
important role in the $\pi N \rightarrow \pi \pi N$ reaction close to
threshold \cite{Oset85,Bernard95}, which must be brought under control
in order to make predictions about the $\pi \pi $ scattering length, $etc$.

The proposed experiments exciting the Roper with an isoscalar source will
bring new information about this resonance, its decay and its coupling to
different hadronic components and will pose new challenge to models of
this resonance.

\vspace{1truecm}

We would like to acknowledge useful discussions with E. Strokovsky.
  One of
us (S. H.) acknowledges the hospitality of Departamento de F\'{\i}sia
Te\'orica, Univ. de Valencia where this work has been done.  This work
has been partly supported by CICYT contract no. AEN 93-1205.

\clearpage

{\bf Figure Caption\\}

\setlength{\baselineskip}{8mm}

\vspace{0.3truecm}

\noindent
{\bf Fig. 1} \quad Calculated cross sections of the target Roper
process \cite{HirenSub} and the projectile $\Delta$ process \cite{Pedro95}
 at $E_{\alpha}= 4.2$ $GeV$
and $\theta_{Lab} = 0.8^{o}$.  The variable $\omega$
is the energy transfer defined as
$\omega = E_{\alpha} - E_{\alpha '} $.  The thick line
indicates the sum of all contributions. Experimental data are taken from
ref. \cite{MorschPC}. Here we used $g ^2_{ \sigma N N^{*}}/ 4 \pi = 1.33,
M^*= 1430MeV, \Gamma^* (s=M^{*2})= 300MeV$.

\noindent
{\bf Fig. 2} \quad Diagrams for the $(\alpha,\alpha')$ reaction which
we consider in this paper.  They are (a) $\Delta$ excitation in
the projectile \cite{Pedro95}, (b) Roper excitation in the
target \cite{HirenSub}.  The $\sigma$ exchange must be
interpreted as an effective interaction in the $T=0$ exchange
channel \cite{HirenSub}.

\noindent
{\bf Fig. 3} \quad Same as Fig. 1.  Here $E_{\alpha}=15$ $GeV$ and
$\theta_{Lab} = 0^{o}$.

\noindent
{\bf Fig. 4} \quad The squared ratio of the $\alpha$ form factor is
plotted as a function of the energy transfer
$\omega = E_{\alpha} - E_{\alpha '} $.  The line (1) indicates the
squared ratio of the form factor of $E_{\alpha}=10$ $GeV, \theta_{Lab}=0^{o}$
case to $E_{\alpha} =4.2$ $GeV, \theta_{Lab}= 0.8^{o}$ case, and the line (2)
the squared ratio of $E_{\alpha}=15$ $GeV, \theta_{Lab}=0^{o}$ to
$E_{\alpha} =4.2$ $GeV, \theta_{Lab}=0.8^{o}$. The form factor $F$ is
defined by eq. (3) in text.
\end{document}